\documentclass[letterpaper,preprintnumbers,prd,twocolumn,nofootinbib,nobibnotes,showpacs]{revtex4}
\usepackage{epsfig}
\usepackage{graphicx}%
\usepackage{dcolumn}
\usepackage{amsmath}

\makeatletter
\def\btt#1{\texttt{\@backslashchar#1}}%
\DeclareRobustCommand\bblash{\btt{\@backslashchar}}%
\makeatother

\begin{document}

\title{A Holographic Dark Energy Model from Ricci Scalar Curvature}

\author{Changjun Gao}
\author{Fengquan Wu}
\author{Xuelei Chen}
\affiliation{The National Astronomical
Observatories, Chinese Academy of Sciences, Beijing, 100012, China}

\author{You-Gen Shen}
\affiliation{Shanghai Astronomical Observatory, Chinese Academy of Sciences,
Shanghai 200030, China}
\affiliation{Joint Institute for Galaxy and Cosmology of SHAO and
  USTC, Shanghai 200030, China}

\date{November 2008}

\begin{abstract}
Motivated by the holographic principle, it has been suggested that
the dark energy density may be inversely proportional to the area of
the event horizon of the Universe. However, such a model would have
a causality problem. In this paper, we propose to replace the future
event horizon area with the inverse of the Ricci scalar curvature.
We show that this model does not only avoid the causality problem
and is phenomenologically viable, but also naturally solves the
coincidence problem of dark energy.
Our analysis of the evolution of density perturbations show that
the matter power spectra and CMB temperature anisotropy is
only slightly affected by such modification.
\end{abstract}
\pacs{98.80.Cq, 98.65.Dx}

\maketitle


\section{Introduction}

Ever since the discovery of the dark energy \cite{Perlmutter:99,
Riess:98}, cosmologists are confronted with two fundamental
problems: (1) the fine tuning problem and (2) the coincidence
problem (see e.g. \cite{weinberg:2000}). The fine tuning problem is the
following: the simplest form of dark energy is the cosmological
constant introduced by Einstein. However, the vacuum
energy in quantum field theory has exactly the same property, and
the estimated size of the vacuum
energy is $\rho\simeq \rho_p$ where $\rho_p \sim m_p^4 $ is the Plank density.
It is greater than the observed value $\rho\simeq 10^{-123}\rho_p$ by
some $123$ orders of magnitude, so extreme fine tuning of the vacuum energy is
required. The coincidence problem is the
following: the density of the dark energy and matter evolves
differently as the Universe expands, yet they are comparable today,
this is an incredibly great coincidence if there is not some internal connection
between the two.

An important advance in the studies of black hole theory and string
theory is the suggestion of the so called holographic principle,
which may provide some clue for solving these problems. It is
realized that in quantum gravity, the entropy of a system scales not
with its volume, but with its surface area $L^2$\cite{holography}.
To see how this could help solve the cosmological constant problems,
we note that in the Einstein equation, $G_{\mu\nu}=8\pi G
T_{\mu\nu}+\Lambda g_{\mu\nu}$, the cosmological constant $\Lambda$
is the inverse of some length squared, $[\Lambda] \sim l^{-2}$, and
to be consistent with observations, $l$ must be of the same order as
the present cosmological scale. It is then proposed
\cite{horova:2000} that an unknown vacuum energy could be present,
and according to the holographic principle its density is
proportional to the Hubble scale $l_H \sim H^{-1}$. In this model
the fine tuning problem is solved as the scale of dark energy is
determined not by Planck length but by cosmological scale, and the
coincidence problem is also all alleviated. Unfortunately,  the
effective equation of state for such vacuum energy is zero and the
Universe is decelerating. Alternatively, the particle horizon size
$l_{PH}=a\int_{0}^{t}{dt}/{a}$ could be used as the length scale
\cite{fischler:1998}. However, as S. Hsu \cite{hsu:2004} and M. Li
\cite{li:2004} pointed out, the equation of state for this dark
energy model is greater than $-1/3$, so it still could not explain
the observed acceleration of the Universe.  In view of this, M. Li
\cite{li:2004} proposed that the future event horizon of the
Universe to be used as the characteristic length $l$. This {\it
holographic dark energy model} and its interacting versions are
successful in fitting the current observations \cite{Zhang:2005}.

However, the underlying origin of the holographic dark energy is
still unknown. Furthermore, the model also has some serious
conceptual problems. As R. Cai \cite{cai:2007} pointed out, an
obvious drawback concerning causality appears in this proposal.
Event horizon is a global concept of space-time. However, the
density of dark energy is a local quantity. Why should a local
quantity be determined by a global one? Also puzzling is that the
present value of dark energy is determined by the future evolution
of the Universe, thus posing a challenge to the concept of
casuality. Furthermore, for a spatially flat
Friedmann-Robertson-Walker Universe, it is well-known that the
future event horizon exists if and if the Universe is accelerating.
So in order to interpret the cosmic acceleration, the holographic
dark energy model itself has presumed the acceleration.

Inspired by the holographic dark energy models, in this paper
 we propose to consider another possibility: the length $l$ is
giving by the average radius of Ricci
scalar curvature, $R^{-1/2}$, so that we have the dark energy
$\rho_X \propto R$. In the following
we shall call this model the {\it Ricci dark energy model}, and
investigate its phenomenological properties.
We find that this model works fairly well in fitting the observational data,
and it could also help us to understand the coincidence problem.
Moreover, in this model the presence of event horizon is not presumed, so
the causality problem is avoided.

In the next section we describe the model and its cosmic expansion
history. Section III is devoted to the study of structure formation
and CMB anisotropy in this model. In Section IV, we summarize our
results, and discuss various problems, including how it solves the
fine-tuning and coincidence problem, while avoids the problem of
causality. We also discuss possible physical mechanism for such a
dark energy. We also show that one could construct a K-essence model
to mimic its behavior in the Appendix. Throughout the paper we adopt
the Planck units, i.e. $c=\hbar=G=1$.

\section{The Model}

The metric of the Friedmann-Robertson-Walker Universe is given by
\begin{equation}
ds^2=-dt^2+a\left(t\right)^2\left(\frac{dr^2}{1-kr^2}+r^2d\theta^2+r^2\sin^2\theta
d\phi^2\right),
\end{equation}
where $k=1,0,-1$ for closed, flat and open geometries respectively. In
this we have adopted the convention of $a_0=1$, where the subscript 0 denotes
the value at present time (zero redshift). The Friedman equation is
\begin{equation}
H^2 = \frac{8\pi}{3} \sum_i \rho_i - \frac{k}{a^2}
\end{equation}
where $H\equiv \dot{a}/a$ is the Hubble parameter, dot denotes the
derivative with respect to the cosmic time $t$, and summation runs
over the non-relativistic matter, radiation and other components.
The Ricci scalar curvature is given by
\begin{equation}
R=-6\left(\dot{H}+2H^2+\frac{k}{a^2}\right),
\end{equation}
We consider a dark energy component,
which arises from unknown physics and is proportional to the inverse of squared
Ricci scalar curvature radius as prescribed by the holographic principle,
\begin{equation}
\label{eq:X}
\rho_X=\frac{3\alpha}{8\pi}\left(\dot{H}+2H^2+\frac{
k}{a^2}\right)=-\frac{\alpha}{16\pi} R,
\end{equation}
where $\alpha$ is a constant to be determined. The factor
$\frac{3}{8\pi}$ before $\alpha$ is for convenience in the following
calculations. The corresponding energy-momentum tensor can be
written as:
\begin{eqnarray}
T_{\mu\nu}=\left(\rho_{X}+p_{X}\right)U_{\mu}U_{\nu}+p_{X}g_{\mu\nu},
\end{eqnarray}
where $U_{\mu}$ is the 4-velocity of the co-moving observer, and $p_{X}$
is the pressure of dark energy.
Setting $x=\ln a$, we can rewrite the Friedmann equation as follows
\begin{eqnarray}
H^2&=&\frac{8\pi}{3}\left[{{\left(\alpha
-1\right)\frac{3k}{8\pi}}}{e^{-2x}}+{\rho_{m}}{e^{-3x}}+{\rho_{r}}{{e^{-4x}}}\right]\nonumber\\
&&+\alpha\left(\frac{1}{2}\frac{dH^2}{dx}+2H^2\right),
\end{eqnarray}
where $\rho_m$ and $\rho_{r}$ term are the contributions of non-relativistic
matter and radiation, respectively.
We introduce the scaled Hubble expansion rate ${\mathfrak h}\equiv H/H_0$,
then the above Friedman equation becomes
\begin{eqnarray}
\label{eq:h}
{\mathfrak h}^2&=&\left(\alpha-1\right)\Omega_{k0} e^{-2x}+\Omega_{m0}
e^{-3x} +\Omega_{r0} e^{-4x}+\nonumber\\
&&\alpha\left(\frac{1}{2}\frac{d{\mathfrak h}^2}{dx}+2{\mathfrak h}^2\right),
\end{eqnarray}
where $\Omega_{k0}$, $\Omega_{m0}$ and $\Omega_{r0}$ are the relative density
of the curvature, non-relativistic matter and radiation
in the present Universe, and the dark energy relative density is
denoted by $\Omega_X$, with
$\Omega_{k0}+\Omega_{m0}+\Omega_{r0}+\Omega_{X0}=1$.
Solving Eq.(\ref{eq:h}), we obtain
\begin{eqnarray}
\label{eq:h-Omega}
{\mathfrak h}^2&=&-\Omega_{k0} e^{-2x}+{\Omega_{m0}}e^{-3x}+\Omega_{r0}e^{-4x}
+\nonumber\\
&&\frac{\alpha}{2-\alpha}\Omega_{m0}e^{-3x}+f_0e^{-\left(4-\frac{2}{\alpha}\right)x},
\end{eqnarray}
where $f_0$ is an integration constant. On the right hand side of
Eq.(\ref{eq:h-Omega}), the last two terms come from the
dark energy,
\begin{equation}
\label{eq:rho_DE}
\rho_X=\frac{\alpha}{2-\alpha}\Omega_{m0}e^{-3x}+f_0e^{-\left(4-\frac{2}{\alpha}\right)x}.
\end{equation}
Thus the Ricci dark energy has one part which evolves like
non-relativistic matter ($\sim e^{-3x}$), and another part which is
slowly increasing with decreasing redshift.

We assume that energy is conserved in such model,
substituting the expression of $\rho_X$ into the conservation
equation of energy,
\begin{equation}
\label{eq:conserve}
p_X=-\rho_X-\frac{1}{3}\frac{d\rho_X}{dx},
\end{equation}
we obtain the pressure of dark energy
\begin{equation}
\label{eq:p_DE}
p_X=-\left(\frac{2}{3\alpha}-\frac{1}{3}\right)f_0e^{-\left(4-\frac{2}{\alpha}\right)x}.
\end{equation}
There are two constants $\alpha$ and $f_0$ to be determined in the
expressions of $\rho_X$ and $p_X$. If the density  $\Omega_X$ and equation of state
$w_{0}\Omega_X$ of the dark energy is known, the value of
$\alpha$ and $f_0$ can be determined using
Eq.~(\ref{eq:rho_DE}) and Eq.~(\ref{eq:p_DE}):
\begin{equation}
\frac{\Omega_{m0}\alpha}{2-\alpha}+f_0=\Omega_{X0},\ \ \ \
-\left(\frac{2}{3\alpha}-\frac{1}{3}\right)f_0=w_{0}\Omega_{X0}.
\end{equation}
We then obtain
\begin{equation}
f_0=\frac{3w_{0}\Omega_{X0}^2}{3w_{0}\Omega_{X0}-\Omega_{m0}},\ \ \ \
\alpha=\frac{2\Omega_{X0}}{\Omega_{m0}+\Omega_{X0}-3w_0\Omega_{X0}}.
\end{equation}

\begin{figure}[htbp]
\begin{center}
\includegraphics[scale=0.8]{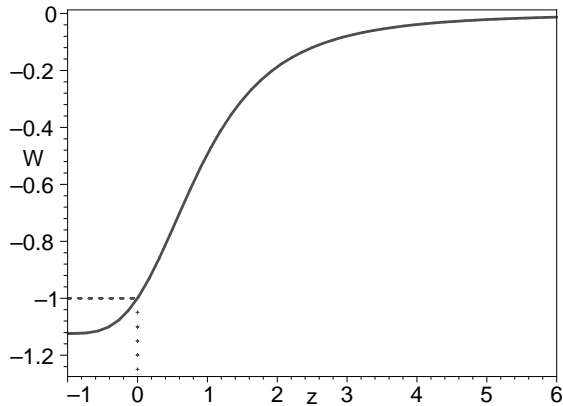}
\caption{\label{fig:w-z} Evolution of the equation of state $w$ for
the Ricci dark energy as a function of redshift $z$.
}
\end{center}
\end{figure}

\begin{figure}[htbp]
\begin{center}
\includegraphics[scale=0.8]{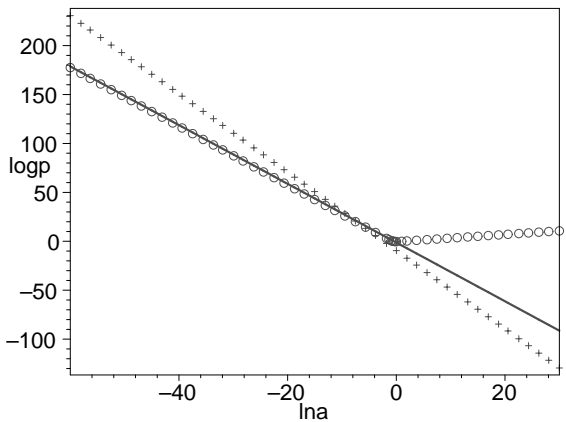}
\caption{\label{fig:lnp-lna} Evolution of the radiation density
(crosses), non-relativistic matter density (solid line) and Ricci
dark energy density (circles) $\log\rho$ as the function of $\ln
a$. }
\end{center}
\end{figure}

As an example, we consider the cosmology model with the following
values of parameters: $\Omega_{k0}=0, \Omega_{m0}=0.27,
\Omega_{r0}=8.1 \cdot 10^{-5}, \Omega_{X0}=0.73,  w_0=-1$, which are
consistent with current observations \cite{spergel:2007}, then we
find $f_0\simeq 0.65$ and $\alpha\simeq 0.46$. We plot the evolution
of the equation of state $w\equiv p_X/\rho_X$ for this model in
Fig.~\ref{fig:w-z}. At high redshifts, the equation of state is
nearly zero, so the Ricci dark energy behaves just like dark matter,
with $\rho_X/\rho_{m} \simeq 0.29$. The equation of state $w$
approaches -1 at $z \sim 0$.  In the distant future, the equation of
state approaches $w=-1.12$, The Universe evolves into the phantom
dominated epoch \cite{caldwell:1999}. For this model, the equation
state crosses -1, so it may be classified as a ``quintom''
\cite{feng: 2004}.

In Fig.\ref{fig:lnp-lna}, we plot the evolution of densities,
$\log \rho$, for radiation (crosses), non-relativistic matter
(solid line) and dark energy (circles) with $\ln a$. Here we
have neglected phase transitions, new degrees of freedoms and transitions from
  non-relativistic to relativistic particles at high temperature,
  etc, which  would not make qualitative difference in the result.
In this model the densities of non-relativistic
matter and dark energy were comparable with each other
in the past Universe, but the acceleration began at low redshift,
so the coincidence problem is solved.
The dark energy component made negligible contribution in the
epoch of radiation dominated Universe, hence the standard Big-Bang
Nucleosynthesis (BBN) model needs no revision.

\begin{figure}[htbp]
\includegraphics[scale=0.8]{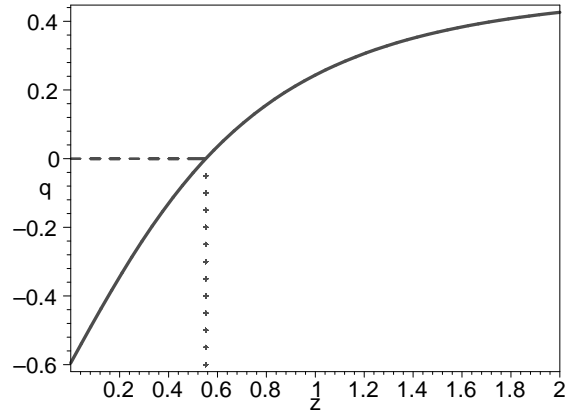}
\caption{\label{fig:q-z}
Evolution of the deceleration parameter with redshift.}
\end{figure}

In Fig.\ref{fig:q-z}, we plotted the evolution of deceleration parameter,
\begin{equation}
q\equiv\frac{1}{2}\left(1+\frac{3p_{tot}}{\rho_{tot}}\right)=\frac{1}{2}+\frac{3p_X}{2\rho_X+2\rho_{m}},
\end{equation}
where $p_{tot}, \rho_{tot}$ denote the total pressure and density of
the Universe respectively. The acceleration of the Universe starts at $z\simeq 0.55$.
For comparison, joint analysis of SNe+CMB data with the $\Lambda$CDM model yields
 $z_T =0.52-0.73$ \cite{alam:2004}.

\begin{figure}[htbp]
\includegraphics[scale=0.8]{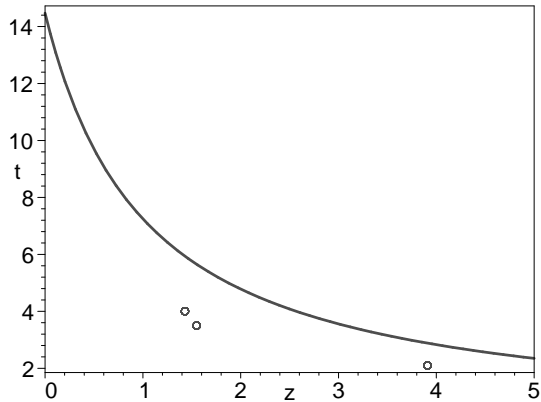}
\caption{\label{fig:t-z}
Age of the Universe with redshift.
Three circles denote the ages of three old
objects, LBDS 53W091 , LBDS 53W069 and APM 08279+5255 .
}
\end{figure}

In Fig.\ref{fig:t-z}, we plotted the evolution of age of the Universe
\begin{equation}
t=\frac{1}{H_0}\int_0^{\frac{1}{1+z}}\frac{dx}{\mathfrak h}.
\end{equation}
Three circles denote the ages of several old objects, LBDS 53W091
($z=1.55$, $t=3.5$ Gyr) \cite{dunlop: 1996}, LBDS 53W069
($z=1.43$,$t=4.0$ Gyr)  \cite{spinrad:1999} and APM 08279+5255
($z=3.91, t=2.1$ Gyr) \cite{jain:2006}. H. Wei and S. N. Zhang
\cite{wei:2007} recently pointed out that the ages of
these three old high redshift objects are inconsistent with
 the holographic dark energy
model of Ref.~\cite{li:2004}.
However, as shown in Fig.4, our model of Ricci dark energy does not
suffer from this age problem.

\section{structure formation}
From Eq.~(\ref{eq:rho_DE}), we see that  if $\alpha \approx 0.46$ which yields
the correct dark energy density and equation of state today,
then the dark energy component behaves much as dust matter
during most of the epoch of matter domination. Thus at earlier time the
matter component is greater than in the best fit $\Lambda$CDM model, the
difference could be as large as about $30{\%}$. An obvious question then is whether
this would affect the growth rate of density perturbation,
and upset the usual structure formation scenario?

\begin{figure}[htbp]
\begin{center}
\includegraphics[scale=0.6]{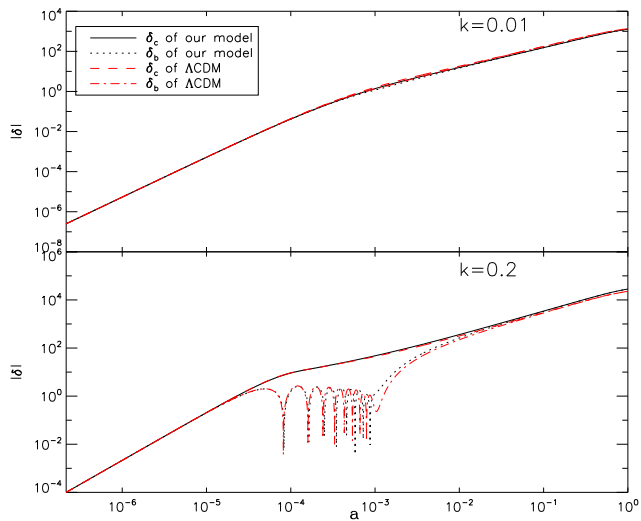}
\caption{\label{fig:delta} Evolution of the density perturbations in
the Ricci dark energy model and $\Lambda$CDM model for
two wavenumbers: k=0.01 (upper panel) and 0.2 h Mpc${}^{-1}$ (lower
panel).  $\delta_c$ and $\delta_b$ are the perturbations
of cold dark matter and baryon respectively. }
\end{center}
\end{figure}

We now investigate this problem with numerical calculations by
using a modified version of the Boltzmann
code \textit{CAMB}\cite{camb}.  In Figure \ref{fig:delta}, we
compare the evolution of density perturbations
in our model with that in the $\Lambda$CDM model.
The parameters as discussed earlier are
 $\Omega_{k0}=0,\Omega_{m0}=0.27 (\Omega_{b}=0.04,\Omega_{c}=0.23),
\Omega_{X0}=0.73$, and $w_0=-1$.

Figure \ref{fig:delta} shows that for the large scale perturbations
(e.g. k=0.01 h Mpc${}^{-1}$ modes) the differences between our model and
the $\Lambda$CDM model are very small, almost
invisible. For the small scale perturbations (e.g. k=0.2 h
Mpc${}^{-1}$ modes), the amplitudes of $\delta_c$ and $\delta_b$ in
our model are slightly (about 20\%) larger  than  these in $\Lambda
\textrm{CDM}$ model, due to the extra dust-like
component in our model.

\begin{figure}[htbp]
\begin{center}
\includegraphics[scale=1.3]{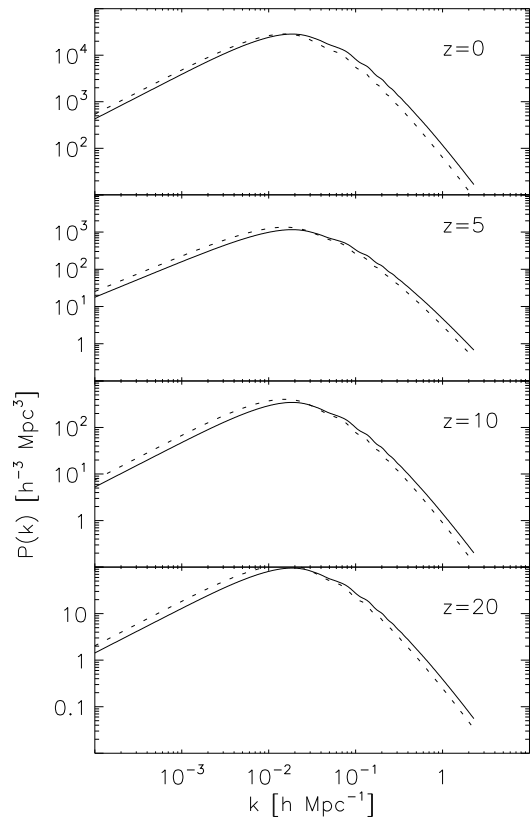}
\caption{\label{fig:spectrum.matter} The matter power spectra at
different redshifts. From top to bottom: z=0, z=5, z=10 and z=20.
The solid and dashed curves represent our model and $\Lambda
\textrm{CDM}$ model respectively. }
\end{center}
\end{figure}

We plot the matter power spectra at different redshifts in
Fig.~\ref{fig:spectrum.matter}. As expected, due to the extra
dust-like component, the matter-radiation equality occurred at
smaller $a_{eq}$, so the turn over in the matter power spectrum are
also at smaller scales. Also, the growth rate of our model differs
from that in $\Lambda$CDM model. However, the deviation of the
shapes of the spectra from $\Lambda$CDM model is not too large, it
is expected that the observation could be fitted well by adjusting
other parameters, such as $\sigma_8$ and $n_S$.

\begin{figure}[htbp]
\begin{center}
\includegraphics[scale=0.5]{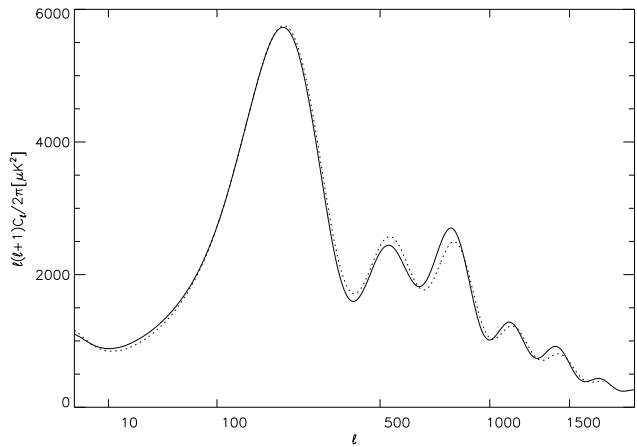}
\caption{\label{fig:spectrum.tt} The theoretical CMB TT spectrum of
our model (solid curve) compared with  $\Lambda \textrm{CDM}$ model
(dashed curve).  We normalize  two amplitudes of spectra to same at
the first peak.}
\end{center}
\end{figure}

We have also calculated the angular spectrum of CMB temperature anisotropy, as
shown in Fig.\ref{fig:spectrum.tt}. It differs from that of $\Lambda$CDM
model mostly at small scales, where the experimental error bars are still very
large, so our model is not in conflict with current CMB
observations. Future high precision observations (e.g. from the Planck satellite)
may help distinguish our model from the $\Lambda \textrm{CDM}$ model.

\begin{figure}[htbp]
\begin{center}
\includegraphics[scale=0.5]{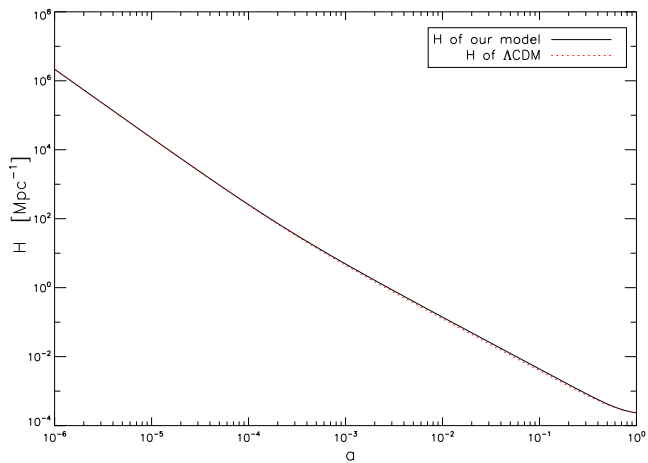}
\caption{\label{fig:H} Evolution of the Hubble parameter of our
model and the $\Lambda$CDM model.}
\end{center}
\end{figure}

At first sight, it may be a little surprising that such large extra
contribution to the dust-like matter produce only such small
difference. To better understand this, we plot the time evolution of
the Hubble parameter in our model and the $\Lambda$CDM model in
Fig.~\ref{fig:H}. As can be seen from the figure, during the
radiation dominated era, the Hubble rates are almost identical--for
the matter contribution is dynamically negligible then. In the
matter dominated era, there is some difference, but the largest
difference is only 10\% at $a=10^{-2}$. Then as dark energy
dominates, again the difference between the two models become very
small. We know that during the matter dominated epoch, the growth
rate of density fluctuation is proportional to scale factor $a$,
since the difference in expansion history is small, so the
modification to structure formation history is also slight.

\section{Discussion}

We have shown that if we replace the future event horizon
in the holographic dark energy model with the Ricci
scalar curvature radius, i.e., $\rho_X\propto R$, then the resulting
Ricci dark energy model is viable phenomenologically. The expansion history of the
Universe in this model is similar to that of $\Lambda$CDM at late times which
has probed by supernovae observation. The model is also
free of the age problem, which plagues
the holographic dark energy model. We have only illustrated our
model with one simple example. For different cosmological parameters,
the model differs slightly. The best fit parameters and their allowed ranges
will be investigated in future work.

We have also studied the structure formation in our model. The
result shows that despite of the dust-like contribution from the
dark energy component during the epoch of radiation domination, the
matter power spectrum and CMB angular spectrum deviates from the
$\Lambda$CDM case only slightly. In this paper we have only
considered the large scale effect of $\rho_X \propto R$. In fact,
due to inhomogeneities, $R$ has spatial fluctuations. The effect of
such fluctuations will also be dealt with in our future study.

Our model has avoided the casuality problem of holographic dark energy, because the
dark energy is determined by the locally determined Ricci scalar curvature, not the
future event horizon. Moreover,
in our model the fine tuning problem is avoided, because the dark
energy is not associated Planck or other high energy physics scale,
but with the size of space-time curvature. Interestingly, the
coincidence problem is also solved in this model: as the dark energy is
proportional to $R$, it is relatively small during radiation dominated era\footnote{The actual
value of the $R$ is not necessarily small during radiation domination, but its value relative
to the Hubble rate is small so its influence on dynamics of expansion can be neglected. See
Ref.\cite{CK:99} for discussion on this point.}. Once entering the matter dominated era,
$R^{1/2}$ become comparable to $H$, the dark energy also become dynamically significant, and
is always comparable to the size of the non-relativistic
matter by construction. The change from deceleration to acceleration happens near
$a \sim 1$ with very plausible model parameters.

While this model is phenomenologically viable and also solves the fine tuning problem and
coincidence problem, one may question whether there is any physical
mechanism or reasoning for which such a model could be motivated. As there is so much
things unknown and uncertain in quantum gravity at present,
it is difficult to make a very definite answer to this question. Rather, we take the view
that the phenomenological success of our model may motivate theorists to look for such
mechanism in their investigations of quantum gravity. However, we
do try to answer this problem from the following two perspectives.

First, we considered the construction of a K-essence model \cite{k-essence:00}which reproduce
the behavior of our model. The K-essence is a ``designer fluid'' with adjustable equation
of state, which was first proposed as a model of dark energy to solve the coincidence problem.
Moreover, the K-essence equation of state can also cross -1 \cite{vik:2005}.
As such, it is not surprising that for {\it each} of our model with a given set of model
parameters, one can build a model of K-essence to mimic its behavior.
We describe the construction of such a model in the Appendix.
Interestingly,  for small $\alpha$ the K-essence action reduces to the
recently proposed Cuscuton action \cite{cuscuton:2006}.
We find that the speed of sound in this model is positive, so it is classically stable.
However, the fundamental dynamics of the K-essence model are of course
not the same as that of our model, so the K-essence model can not realize the
general mechanism of self-adjustment of our model.

Second, we may also consider some heuristic arguments based on conjectures about
quantum gravity effects, though of course the argument is unavoidably vague and speculative
before we have a better understanding of quantum gravity. An an example,
one may suppose that the dark energy raises due to fluctuation in quantum gravity.
The gravitational action is given by
$$S=\frac{1}{16\pi}\int d^4x\sqrt{-g}R.$$
In classical evolution, $\delta S=0$. However, in quantum gravity,
at some minimal scale, it has been suggested \cite{gar:1994} that
this action might be small but non-zero due to quantum fluctuation,
$\Delta S \simeq 1$ within the space-time volume $\Delta V \Delta
t$, so that
$$\Delta S = \frac{R}{16\pi} \Delta V \Delta t \sim 1.$$
The density fluctuation in such small
region is given by the energy-time uncertainty relation, $\Delta E \Delta t \sim 1$, so
$$\rho_X \sim \frac{\Delta E}{\Delta V} \propto R,$$
which is what we have taken as the hypothesis of our model.

Finally, we point out that although we have been motivated by the holographic
principle when constructing our model, the Ricci dark energy does not necessarily
have to be connected with the holographic principle. The dark energy associated with
the Ricci scalar may also arise for other physical reasons. The main point of this paper
is to show that dark energy of such property would be successful.

\appendix
\section{K-essence construction}
Here we show that for each of our model, one can also construct a K-essence
model \cite{k-essence:00} which mimics its behavior.

As shown in the section II, the equation of state of the Ricci scalar dark energy
may cross over $-1$. This can be realized
with K-essence field with Lagrangian density $p~(\phi,X)$ \cite{vik:2005}
where $\phi$ a scalar field.

Now let us consider the Lagrangian density of a K-essence field
$p~(\phi,Y)$, where $Y=-\frac{1}{2}\partial
_{\mu}\phi\partial^{\mu}\phi$ is the kinetic energy term. We note
that since we chose the signature of the metric as (-1, +1, +1,
+1,), $Y$ is positive. For simplification, here we consider the pure
K-essence with the Lagrangian density
\begin{eqnarray}
 \label{eq:lag}
 p=p\left(Y\right)\;.
\end{eqnarray}
Identifying the energy momentum tensor of the scalar field with that of a perfect
fluid, we can easily derive the energy density and pressure:
\begin{eqnarray}
 \label{eq:den-pre}
 \rho_Y =2Y\frac{dp}{dY}-p\;,\ \ \ p_Y=p\;.
\end{eqnarray}
Put $\rho_X=\rho_Y$ and $p_X=p_Y$, we have
\begin{eqnarray}
 \label{eq:den1-pre1}
 2Y\frac{dp}{dY}-p&=&\frac{\alpha}{2-\alpha}\Omega_{m0}e^{-3x}+f_0e^{-\left(4-\frac{2}{\alpha}\right)x}\;,\nonumber\\
 \ \ \ p&=&-\left(\frac{2}{3\alpha}-\frac{1}{3}\right)f_0e^{-\left(4-\frac{2}{\alpha}\right)x}\;.
\end{eqnarray}
Solving Eq.(\ref{eq:den1-pre1}), we find a remarkably simple
form for the action of the pure K-essence field
\begin{eqnarray}
 \label{eq:action1}
 S_{K}&=&\frac{1}{{16\pi}}\int d^4x\sqrt{-g}{p}\nonumber\\ &\propto&\frac{1}{{16\pi}}\int d^4x\sqrt{-g}
 \left(\sqrt{-\frac{1}{2}\partial _{\mu}\phi\partial^{\mu}\phi}-V_0\right)^{\frac{2\left(1-2\alpha\right)}{2-\alpha}} \;,
 \end{eqnarray}
where $V_0$ is a positive integration constant. We note that when $\alpha\ll1$, this action
reduces to the Cuscuton action with a constant scalar potential which were
proposed by N. Afsgordi, D. J. H. Chung and G. Geshnizjani recently \cite{cuscuton:2006}.
The energy density of this model is easily shown to be positive using Eq.~(\ref{eq:den-pre}).
The solution is stable with respect to small perturbation to the Cauchy data if the effective
sound speed is positive. The sound speed of K-essence is given by \cite{k-essence:00}
\begin{eqnarray}
 \label{eq:speed}
 v_s^2\equiv\frac{p_{Y,Y}}{\rho_{Y,Y}}=\frac{\left(\sqrt{Y}-V_0\right)\left(\alpha-2\right)}{3\alpha\sqrt{Y}} \;.
 \end{eqnarray}
Since $\alpha\simeq 0.46$, so if $\sqrt{Y}\leq V_0$,
we will have $v_s\geq 0$. Thus we have a viable K-essence model which mimics the behavior of our
model.


\acknowledgments
We thank  professors Miao Li, Ronggen Cai, Chongming Xu, Pengjie Zhang
and Andrew R. Liddle for helpful discussions. This work is supported by the
National Science Foundation of China under the Distinguished Young
Scholar Grant 10525314, the Key Project Grant 10533010, and Grant
10575004; by the Chinese Academy of Sciences under grant
KJCX3-SYW-N2; and by the Ministry of Science and Technology under
the National Basic Sciences Program (973) under grant 2007CB815401.

\newcommand\AL[3]{~Astron. Lett.{\bf ~#1}, #2~ (#3)}
\newcommand\AP[3]{~Astropart. Phys.{\bf ~#1}, #2~ (#3)}
\newcommand\AJ[3]{~Astron. J.{\bf ~#1}, #2~(#3)}
\newcommand\APJ[3]{~Astrophys. J.{\bf ~#1}, #2~ (#3)}
\newcommand\APJL[3]{~Astrophys. J. Lett. {\bf ~#1}, L#2~(#3)}
\newcommand\APJS[3]{~Astrophys. J. Suppl. Ser.{\bf ~#1}, #2~(#3)}
\newcommand\JCAP[3]{~JCAP. {\bf ~#1}, #2~ (#3)}
\newcommand\LRR[3]{~Living Rev. Relativity. {\bf ~#1}, #2~ (#3)}
\newcommand\MNRAS[3]{~Mon. Not. R. Astron. Soc.{\bf ~#1}, #2~(#3)}
\newcommand\MNRASL[3]{~Mon. Not. R. Astron. Soc.{\bf ~#1}, L#2~(#3)}
\newcommand\NPB[3]{~Nucl. Phys. B{\bf ~#1}, #2~(#3)}
\newcommand\PLB[3]{~Phys. Lett. B{\bf ~#1}, #2~(#3)}
\newcommand\PRL[3]{~Phys. Rev. Lett.{\bf ~#1}, #2~(#3)}
\newcommand\PR[3]{~Phys. Rep.{\bf ~#1}, #2~(#3)}
\newcommand\PRD[3]{~Phys. Rev. D{\bf ~#1}, #2~(#3)}
\newcommand\RMP[3]{~Rev. Mod. Phys.{\bf ~#1}, #2~(#3)}
\newcommand\SJNP[3]{~Sov. J. Nucl. Phys.{\bf ~#1}, #2~(#3)}
\newcommand\ZPC[3]{~Z. Phys. C{\bf ~#1}, #2~(#3)}



\begin{thebibliography}{99}


\bibitem{Perlmutter:99} S. Perlmutter et al., \APJ{517}{565}{1999}

\bibitem{Riess:98} A. G. Riess et al., \AJ{116}{1009}{1998}

\bibitem{weinberg:2000} S. Weinberg, astro-ph/0005265.

\bibitem{holography} see e.g. G. t' Hooft, gr-qc/9310026; hep-th/0003004;
L. Susskind, J. Math. Phys. 36, 6377 (1995) (hep-th/9409089);
  J. M. Maldacena, Adv. Theor. Math. Phys. 2, 231 (1998); W. Fischler
  and L. Susskind, hep-th/9806039; R. Bousso, \RMP{74}{825}{2000}.

\bibitem{horova:2000} A. Cohen, D. Kaplan and A. Nelson, Phys. Rev. Lett. 82 (1999) 4971;
P. Horava and D. Minic, hep-th/hep-th/0001145, Phys.Rev.Lett. 85 (2000) 1610; S.
Thomas, Phys. Rev. Lett. 89 (2002) 081301.

\bibitem{fischler:1998} W. Fischler and L. Susskind, hep-th/9806039; R. Bousso, JHEP 9907 (1999) 004.

\bibitem{hsu:2004} S. D. H. Hsu, hep-th/0403052.

\bibitem{li:2004} M. Li, Phys.Lett. B603 (2004) 1.

\bibitem{Zhang:2005} R. Horvat, Phys. Rev. D70 (2004) 0873001;
D. Pavon and W. Zimdahl, Phys. Lett. B628 (2005) 206; B. Guberina,
R. Horvat and H. Nikolic, JCAP 0701 (2007) 012; X. Zhang, F. Q. Wu,
Phys. Rev. D 72 (2005) 043524; Z. Chang, F.Q. Wu, X. Zhang,
Phys.Lett.B 633 (2006) 14; X. Zhang, F. Q. Wu, Phys.Rev.D76 (2007)
023502; X. Zhang, Int.J.Mod.Phys.D 14 (2005) 1597; M. R. Setare, J.
Zhang, X. Zhang, JCAP 0703 (2007) 007; J. Zhang, X. Zhang, H. Liu,
Phys.Lett.B659 (2008) 26; X. Zhang, Phys.Lett.B 648 (2007) 1; X.
Zhang Phys. Rev. D74 (2006) 103505,2006; J. Zhang, X. Zhang, H. Liu
Phys.Lett.B651 (2007) 84; Y. Ma, X. Zhang, Phys.Lett.B 661 (2008)
239; J. Zhang, X. Zhang, H. Liu, Eur.Phys.J.C52 (2007) 693; L. Xu,
W. Li£¬ J. Lu£¬ arXiv: 0810.4730 (to appear in MPLA).

\bibitem{cai:2007} R. G. Cai, arXiv:0707.4049v3; I. P. Neupane, Phys. Rev. D 76, 123006
(2007); I. P. Neupane, arXiv: 0708.2910.

\bibitem{spergel:2007} D. N. Spergel et al., \APJS{170}{377}{2007}; Morad
Amarzguioui, O. Elgaroy, David F. Mota and T.
Multamaki,\APJ{454}{707}{2006}.


\bibitem{caldwell:1999} R. R. Caldwell, Phys Lett. B 545 (2002)
23.

\bibitem{feng: 2004}B. Feng, X. L. Wang and X. M. Zhang, Phys. Lett. B 607, 35 (2005) [astro-ph/0404224];
B. Feng, M. Li, Y. S. Piao and X. M. Zhang, Phys. Lett. B 634, 101 (2006) [astro-ph/0407432];

\bibitem{alam:2004} U. Alam, V. Sahni and A. A. Starobinsky, JCAP {\bf 0406}, 008
(2004); Z. H. Zhu, M. K. Fujimoto and X. T. He, Astrophys. J. {\bf
603}, 365 (2004)


\bibitem{dunlop: 1996}J. Dunlop et al., Nature 381, 581 (1996).

\bibitem{spinrad:1999} H. Spinrad et al., Astrophys. J. 484, 581 (1999).

\bibitem{jain:2006} D. Jain and A. Dev, Phys. Lett. B 633, 436 (2006)
[astro-ph/0509212].

 \bibitem{wei:2007} H. Wei and S. N. Zhang, Phys. Rev. D 76, 063003 (2007)


\bibitem{camb} A. Lewis and A. Challinor, http://camb.info/.

\bibitem{Komatsu:2008hk}
  E.~Komatsu {\it et al.}  [WMAP Collaboration],
  arXiv:0803.0547 [astro-ph].

\bibitem{CK:99} X. Chen and M. Kamionkowski, Phys. Rev. D 60 (1999) 104036.



\bibitem{k-essence:00}C. Armendariz-Picon, V. F. Mukhanov and P. J. Steinhardt, Phys. Rev. Lett. 85, 4438 (2000)
[astro-ph/0004134];
C. Armendariz-Picon, V. F. Mukhanov and P. J. Steinhardt, Phys. Rev. D 63, 103510 (2001)
[astro-ph/0006373]; 







\bibitem{vik:2005} A. Vikman, Phys. Rev. D 71, 023515 (2005) [astro-ph/0407107];
S. Tsujikawa, Phys. Rev. D 72, 083512 (2005) [astro-ph/0508542];
E. J. Copeland, M. Sami and S. Tsujikawa, hep-th/0603057.


\bibitem{cuscuton:2006}N. Afshordi, D. J. H. Chung, and G. Geshnizjani, Phys.
Rev. D75 083513 (2007), hep-th/0609150;  G. Robbers, N. Afshordi, M. Doran, Phys. Rev. Lett. 100 (2008) 111101.











\bibitem{gar:1994}L. J. Garay, Int. J. Mod. Phys. A10 (1995) 145-166.
\end{thebibliography}
\end{document}